\documentclass[traditabstract]{aa}\usepackage{graphicx}
\usepackage{txfonts}
\usepackage{layouts}
\usepackage{amssymb}
\usepackage[]{natbib}
\usepackage[colorlinks=true, linkcolor=blue, citecolor=blue,
urlcolor=blue]{hyperref}
\usepackage{longtable}

\begin{document}
\title{Kepler-210: An active star with at least two planets.}
\author{P. Ioannidis $^{1}$, J.H.M.M. Schmitt $^{1}$, Ch. Avdellidou $^{2}$,
C. von Essen $^{1}$ \and E.
Agol $^{3}$}
\institute{$^{1}$Hamburger Sternwarte, Universit\"at Hamburg, Gojenbergsweg 112,
21029
Hamburg, Germany\\
$^{2}$ Centre for Astrophysics and Planetary Science, School of Physical
Sciences (SEPnet), The University of Kent, Canterbury,
CT2 7NH, UK\\
$^{3}$ Dept. of Astronomy, Box 351580, University of Washington,
Seattle, WA 98195\\
\email{pioannidis@hs.uni-hamburg.de}}
\date{Accepted January 31, 2014}

\abstract{ We report the detection and characterization of two short-period,
Neptune-sized planets around the active host star Kepler-210.  The host star's
parameters derived from those planets are (a) mutually inconsistent and (b) do
not conform to the expected host star parameters.  We furthermore report the
detection of transit timing variations (TTVs) in the O-C diagrams for both
planets. We explore various scenarios that explain and resolve those
discrepancies.  A simple scenario consistent with all data appears to be one
that attributes substantial eccentricities to the inner short-period planets
and that interprets the TTVs as due to the action of another, somewhat
longer period planet. To substantiate our suggestions, we present the results
of N-body simulations that modeled the TTVs and that checked the stability of 
the Kepler-210 system.}

\titlerunning {Kepler-210}
\maketitle

\section{Introduction}

Since the launch of the {\it Kepler} Mission in 2009, a large number
of planetary candidates has been found using the transit method in the high
precision  {\it Kepler} light curves.
Specifically, 2321 planetary candidates in 1790 hosts stars have been reported,
from 
which about one third are actually hosted in multiple systems
\citep{2012arXiv1202.5852B}. 
The majority of these  {\it Kepler} planetary candidates are expected to be
real 
planets \citep{2012ApJ...750..112L} and therefore those stars present an
excellent
opportunity for a more detailed study and characterization through the method of
transit timing variations (TTVs). 
Ever since the first proposals of the method by
\cite{2005MNRAS.359..567A} and \cite{2005Sci...307.1288H}, TTVs have been
widely used to search for smaller, otherwise undetectable planets 
in systems containing already confirmed planets.  
In multiple systems this
method can be applied in order to confirm the physical validity of the system 
along with a rough estimate of the components' mass, which can otherwise be
obtained only through radial velocity data. 
For the {\it Kepler}  candidates the
{\it Kepler} team has carried out and reported this kind of analysis for 41 
extrasolar planet systems. For the last announcement of the series see
\cite{2013MNRAS.428.1077S}. 

In this paper we present our in-depth analysis and results for a particular
system, Kepler-210 (=KOI-676), which was previously identified and
listed as a planet host candidate in the catalog by \cite{2011ApJ...736...19B}. 
The specific characteristics of Kepler-210 that enticed us to perform a
detailed 
study of this candidate system were the high activity of its host star coupled
with the fact that the system harbors two transiting planets, which we  
validate using spectral and TTVs analysis, as well as stability tests.

The plan of our paper is as follows. In the first section we
describe the methods used to determine the stellar and planetary properties. 
In the second section we discuss various scenarios to explain the detected
discrepancies in the orbital elements of the planets. Furthermore, we describe
the results of our TTVs analysis for both planets, and finally, we summarize
with
what we 
believe is the most probable scenario.

\begin{figure}[tp]
 \includegraphics[width=\linewidth]{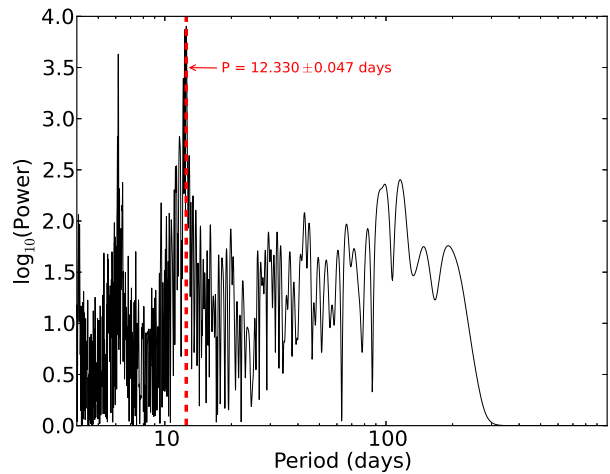}
 \caption{The Lomb-Scargle Periodogram for the raw lightcurve as
logarithmic power (on y axis) vs. period. A polynomial fit was applied to remove
systematics related to the rotation of the telescope.}\label{lsstar}
\vspace{-0.4cm}
\end{figure}

\begin{figure*}[htp]
\label{stellaract}
 \includegraphics[width=\textwidth]{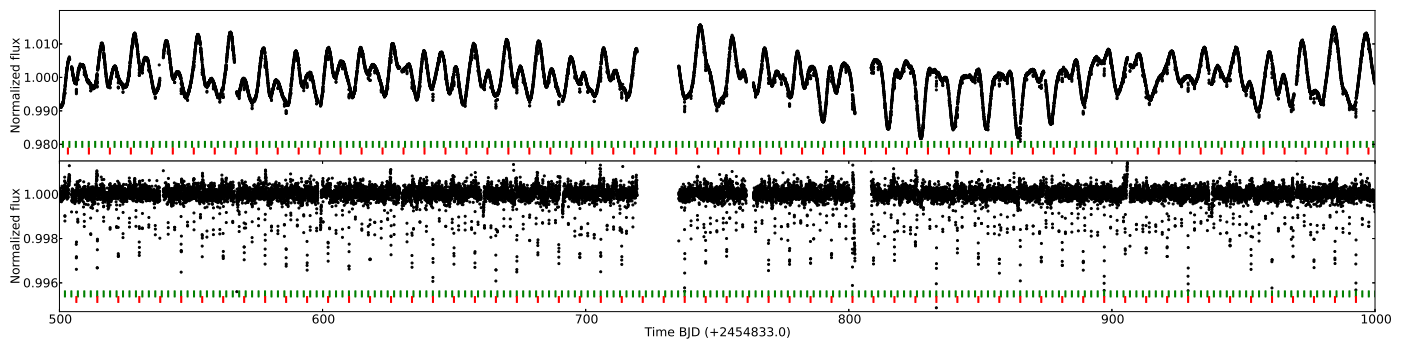}
 \caption{A fraction of the Kepler-210 lightcurve. Top: Part of the raw
lightcurve demonstrating the activity of Kepler-210. Bottom: The
transits of planet a (green) and b (red) for that particular time, with the
stellar activity removed by using the kepflatten routine of the
Pyke package for pyraf.}\label{stellaract}
\end{figure*}

\section{Data analysis}

\subsection{Stellar activity}

\hspace{3 ex}The {\it Kepler} data of Kepler-210 were obtained from the STDADS
archive and contain the data recorded in the quarters Q1 to Q12. 
To achieve better temporal coverage we used both long and short cadence data. 
We decided to use the SAP data for our analysis to avoid
any artifacts introduced by the use of SAP\_PDC data, given the obvious
complexity of the Kepler-210 lightcurve (cf., Figure\ref{stellaract}).
Clearly, the host star of the Kepler-210 system is of particular interest
by itself.  

To give an impression of the activity of Kepler-210, a part of the overall {\it
Kepler} 
lightcurve of Kepler-210 covering 500 days is shown in Fig.\ref{stellaract}.
Peak-to-peak variations on the order of 2\% can easily be identified on 
time scales of a few days, and in addition, variations on longer time scales
are also visible.  
To assess the dominant time scales of variability we computed
a Lomb-Scargle periodogram  over the full available data set, which we show in
Fig.
\ref{lsstar}.  Two peaks are clearly observable in the resulting
periodogram. We interpret the most significant peak at 12.33 $\pm$ 0.15 days as
the rotation period of the
star, while the second, smaller peak at 6.15 $\pm$ 0.047 days  is interpreted as
an alias from the 12.33 day rotation period.  We further note that both peaks
are quite broad, with significant power residing at frequencies near the peak
frequency.

\begin{figure}[bp]
\includegraphics[width=\linewidth]{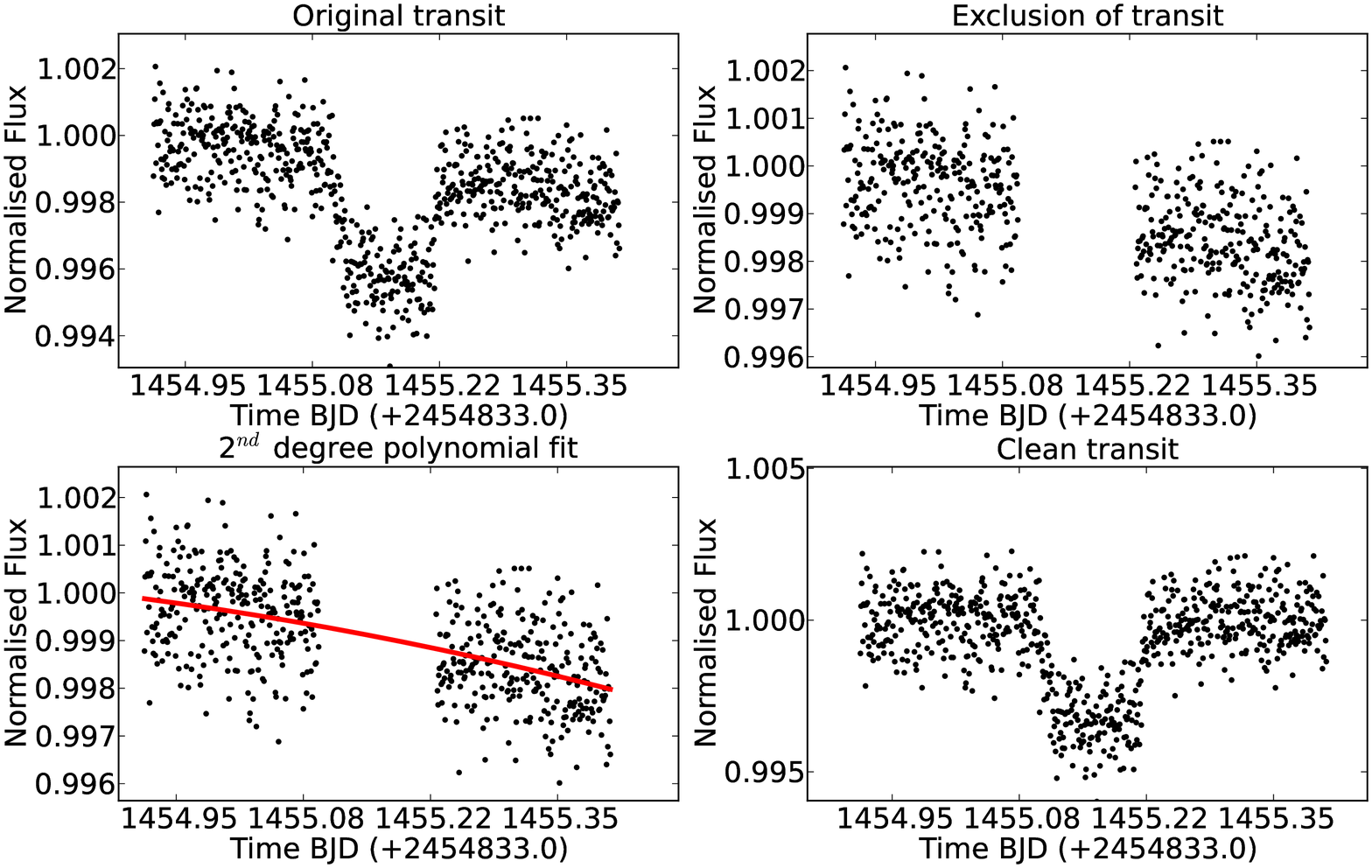}
\caption{Demonstration of stellar activity removal procedure; see text for
details.}\label{deact}
\end{figure}

\subsection{Data Preparation}

\hspace{3 ex} For our transit and TTVs analysis we must remove 
all effects of stellar activity as much as possible.  
In order to rectify the {\it Kepler} lightcurve of Kepler-210 we proceed as
follows:
For each transit of each planet we select some part of the lightcurve centered
at the estimated mid-transit time, including data points before and after
ingress and egress, respectively (see Fig.\ref{deact}, upper left). 
The obvious transit points are then removed (Fig.\ref{deact}, upper right) and 
a second order polynomial fit is applied to the remaining data points 
(see Fig.\ref{deact} lower left). 
Finally, all selected data points including those obtained during transit
are divided by the result of the polynomial fit (see Fig.\ref{deact} lower
right) and we obtain a rectified light curve, normalized to unity for the data
prior to
the first and after the fourth contact. 
In this fashion we prepare the transit data for both planets for an application
of our transit model fit; we consider only transits by one planet and exclude
any
simultaneous transits from our analysis. 

\newcommand{\mc}[3]{\multicolumn{#1}{#2}{#3}}
\begin{table}[tp]
\caption{Stellar parameters of Kepler-210 taken from
\cite{2012arXiv1202.5852B}. The limb-darkening coefficients where calculated 
for those values from \cite{2012yCat..35469014C} for those parameters.}
\begin{center}
\begin{tabular}{lr}
\hline
\hline
\mc{2}{c}{ }\\[-6pt]
\mc{2}{c}{Kepler-210}\\
\hline
&\\[-6pt]
KIC-ID & 7447200\\
KOI-ID & 676\\
T$_{eff}$ & 4300 K\\
logg & 4.55\\
R$_{\star}$ & 0.69 R$_{\odot}$\\
M$_{\star}$ & 0.63 M$_{\odot}$\\
linear LD & 0.7181\\
quadratic LD & 0.0443\\
\hline
\end{tabular}
\end{center}
\label{stellarkeys}
\end{table}

\begin{table}[bp]\label{planetkeys}
\caption{MCMC analysis transit model fit results and their ''1 $\sigma$''
errors.}
\begin{center}
\begin{tabular}{lll}
\hline
\hline
\\[-6pt]
Planet & b & c\\
\hline
&&\\[-6pt]
Period (d) & 2.4532 $\pm$ 0.0007& 7.9725 $\pm$ 0.0014\\
T$_{0}$ (LC) & 134.0952 $\pm$ 0.0002 & 131.7200 $\pm$ 0.0002 \\
R$_{pl}$/R$_{\star}$ & 0.0498 $\pm$ 0.0004 & 0.0635 $\pm$ 0.0006 \\
a/R$_{\star}$ & 4.429 $\pm$ 0.077 & 11.566 $\pm$ 0.323 \\
i &  77.86 $\pm$ 0.26 & 85.73 $\pm$ 0.16 \\
b & 0.931 $\pm$ 0.038 & 0.861 $\pm$ 0.065 \\
transits & 431 lc / 243 sc & 97 lc / 51 sc \\
\hline
\end{tabular}
\end{center}
\label{planetkeys}
\end{table}

\subsection{Model fitting}

\hspace{3 ex}
The effects of the host star's stellar activity are clearly visible also 
in the transit light curves, which are twofold: (see Figure \ref{stellaract}):  
Star spots occulted by the planet on the one hand lead to bumps in the light
curve as shown by
\cite{2009A&A...504..561W}, while star spots on the unocculted face of the star 
on the other hand lead to variable transit depths.  Thus, both
effects increase the dispersion of the transit data and the incorrect
normalization
leads to incorrect stellar and planetary parameters.
In order to minimize the effects of stellar activity we therefore decided to use
transits, following the following two rules:

\begin{enumerate}

\item We select those transits occurring close to maximum flux in the activity
modulations, which corresponds to smaller spot coverage of the stellar surface

\item We model each transit separately, isolated the transit data
and measure the $\chi^2$-test statistics.  Only the transits with acceptable
fits are selected.

\end{enumerate}

Thus, for the transit model fit of the inner planet 89 transits were
used, while for the outer planet a total of 10; we rejected
154 transits by the inner planet and 41 transits by the outer planet for
this analysis.
To determine the best fit model we used the analytical transit light curve model
by \cite{2002ApJ...580L.171M} and Markov-Chain Monte-Carlo
sampling\footnote{ http://www.hs.uni-hamburg.de/DE/Ins/Per/Czesla/PyA/PyA} for
the computation of the fit parameters and their errors.  In this process we used
the limb darkening coefficients
for the model as calculated by integrating the values of the Claret catalog 
\citep{2012yCat..35469014C} for the parent star's nominal parameters
$T_{eff}$, $logg$ and $[Fe/H]$ listed in Table \ref{stellarkeys} . 
\subsection{Fit results} \label{ftres}

\begin{figure}[tp]
 \includegraphics[width=\columnwidth]{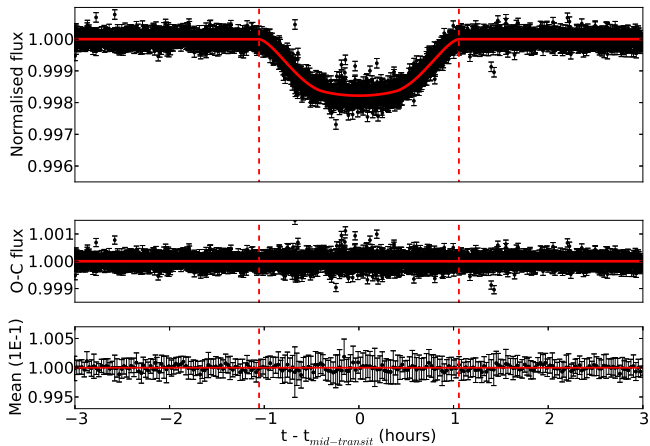}
 \caption{The folded lightcurve of planet b, using the total number of
transits and the best fit model. The lower diagram shows the model residuals
binned, which in this case is less
obvious than in case of planet c (see Fig. \ref{tranmodc1}.}\label{tranmodb1}
\end{figure}

\begin{figure}[tp]
 \includegraphics[width=\columnwidth]{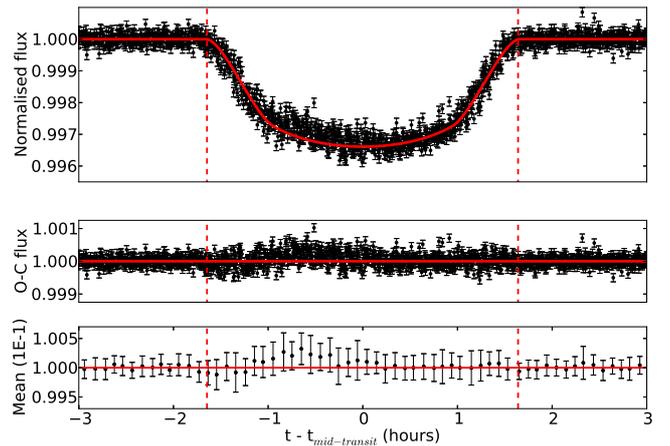}
 \caption{The folded lightcurve of planet c, using the total number of transits
and the best fit model. The
transit profile is affected by spots
during and after ingress.}\label{tranmodc1}
\end{figure}

\hspace{3 ex} In Figs.\ref{tranmodb1} and \ref{tranmodc1} we show the full
sample of the derived
mean normalized {\it Kepler} transit light curves and 
our best fit model (red line; upper panel), the fit residuals for all
data points (middle panel) and the mean values of the individual residuals as
well as for blocks of
twenty adjacent phase points (lower panel). The physical parameters derived from
our transit analysis are listed in Table ~\ref{planetkeys}; note that
the inner, shorter period planet is smaller than the outer planet.

\section{Transit Timing Variations} 
\label{sec:ttvs}

\subsection{Timing variation analysis}\label{subsec:ttvs1}

\begin{figure}[bp]
 \includegraphics[width=\columnwidth]{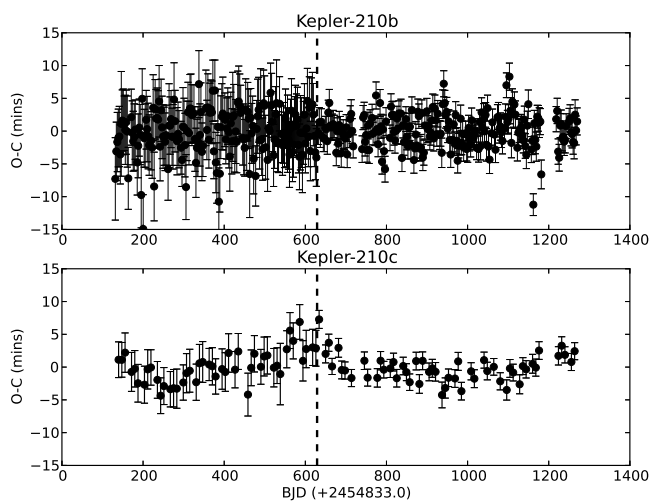}
 \caption{The O-C diagrams of planet b (upper panel) and 
c (lower panel), along with the best fit models. The dashed line discriminates
the long from the short cadence data.}\label{ocdiag} 
\end{figure}

\hspace{3 ex}After completion of the transit model fit procedure and the
determination of the global parameters for each planet, 
we recalculated the mid-transit times for all transits of each planet, i.e., 
also those that had been rejected for the best model fit.
To that end we reapplied the MCMC fitting algorithm for every transit
separately, keeping all model parameters fixed except for the individual
mid-transit
times $t_{MT,i} \; i = 1,N$.  From the observed mid-transit times $t_{MT,i}$,
their errors 
$\sigma_{i}$ and the integer transit epochs $N_{i}$ 
we derived a mean period $P$ and time reference $T_{0}$, by minimizing the
expression 
\begin{eqnarray}
\label{eph}
\chi^{2} = \sum^{N}_{i=1}{(t_{MT,i}-t_{calc,i})^{2}\over\sigma^{2}_{i}} &=& 
\sum^{N}_{i=1}{(t_{MT,i}-P\ {N_i}-T_{o})^{2}\over\sigma^{2}_{i}} \Rightarrow \\
\chi^{2} &=& \sum^{N}_{i=1}{\left( { OC_i\over\sigma^{2}_{i}} \right)}^{2}
\nonumber
\end{eqnarray}
with respect to $P$ and $T_{0}$; the thus resulting O-C diagrams for planets b
and c are shown in Fig \ref{ocdiag}; the appropriate data for the creation of
those diagrams are available in electronic form
at the CDS via anonymous ftp to cdsarc.u-strasbg.fr (130.79.128.5)
or via http://cdsweb.u-strasbg.fr/cgi-bin/qcat?J/A+A/ 


\subsection{TTVs results}

\hspace{3 ex} While in the O-C diagram of planet b (Fig. \ref{ocdiag}) some
modulation
is visible, no clear variation is apparent for planet c.  In order to assess to
what extent
TTVs might be caused by stellar activity, as for example the observed anomaly in
the ingress of the transit of planet b (Fig.
\ref{tranmodb1}), we calculated the transit times using two different
approaches,
additional to the one described in \ref{subsec:ttvs1}:

\begin{itemize}
 \item We recalculated the transit times using the method described in
\ref{subsec:ttvs1}, but by excluding the affected areas of the transit phase. 

 \item We used the analysis described by \cite{2009ApJ...704...51C} in order
to remove the spot anomalies, considering them as red noise. 

\end{itemize}

The results in both cases were almost identical, thus we  decided to use for
our 
analysis the transit times which were produced with the simplest method
(section \ref{subsec:ttvs1}). 
In addition we searched for correlations between the timing variations vs. the
transit depth and duration, but in any case no correlations were found. 

In order to quantify the significance of the observed O-C variations, 
we carried out a $\chi^{2}$-analysis on the null hypothesis that there are no
timing variations, on the long and short cadence data separately and on the
joined  long and short cadence data. We also carried out the same analysis by 
averaging the measured O-C values over ten consecutive epochs; the
respective $\chi^{2}$-values and the derived significance levels are
listed in Table \ref{chisqr}.

An inspection of Table \ref{chisqr} shows that the
unbinned long cadence data show no evidence for any non-zero O-C values,
while the short cadence data do;  binning does greatly increase the significance
of the non-zero O-C values. 

\begin{table}[pt]
\small
\caption{{The null hypothesis $\chi^2$ results} with their
P-values for the
long cadence alone (left column) vs the short cadence alone (middle column) and
the
combined long and short cadence data (right column) for unbinned and binned
data. In the square brackets are listed the degrees of freedom. }\label{chisqr}
\begin{center}
\begin{tabular}{rrrr}
\hline
\hline
\\[-8pt]
  & Long Cad. & Short Cad. & L \& S Mix\\
\hline
\\[-6pt]
b &430.6 (0.47) [430]&347.5 ($>$0.99) [242]&447.9 (0.71)\\
c &111.6 (0.77) [96]&138.7 ($>$0.99) [50]&185.2 ($>$0.99)\\
b$_{bin}$&61.2 (0.88) [43]&63.5 ($>$0.99) [24]&47.5 (0.67) \\
c$_{bin}$&51.9 ($>$0.99) [10]&76.1 ($>$0.99) [5]&67.9 ($>$0.99)\\
\hline
\end{tabular}
\end{center}
\end{table}

Since the values of the $\chi^2$-test statistics sensitively depend on the
measurement errors $\sigma_{i}$, we carefully checked the errors of the derived
mid-transit times by using two independent methods and convinced ourselves 
of the internal consistency of our error determination. In addition we checked
that the
derived $\chi^{2}$-values are not produced by a few individual
outliers.  As a result we are confident that the observed TTVs are statistically
significant.

\section{Discussion}

\subsection{Examination of stellar properties}
\label{sect:age}
In order to better determine the stellar parameters of Kepler-210, 
a high resolution  spectrum was acquired, using the CAFE instrument on the 2.2 m
telescope of the Calar Alto Observatory in Spain.  For our
analysis we also used two spectra of Kepler-210, taken from the
CFOP\footnote {https://cfop.ipac.caltech.edu} web page.
We specifically inspected the spectra for a second set of lines
indicating the existence of a close unresolved companion but found
none. For the same purpose we also examined the pixel area around the star using
the Kepler target fits frames, but found again no evidence of any variable star
in the vicinity of the central star that could create a contaminating signal. 

To measure the color index B-V, we observed Kepler-210 together with
with two standard stars (HD 14827 and HD 195919), using the 1.2 m
Oskar-L\"uhning-Telescope (OLT) from Hamburg Observatory and
found B-V$_{Kepler-210}$ = 1.131 $\pm$ 0.064, consistent with the
color derived from CFOP (1.088 $\pm$ 0.037) and other sources;
thus the spectral type of Kepler-210 is in the late K range.

We then proceeded to estimate the stars age using the gyrochronology expression
\ref{age} derived by  \cite{2007ApJ...669.1167B}
\begin{eqnarray}\label{age}
log(t_{gyro}) &=& {1\over n}\{log(P)-log(\alpha)-[\beta \cdot log(B-V-0.4)]\},
\end{eqnarray}
where t is in Myr, B-V is the measured color, P (in days) is the rotational
period, $n = 0.5189$, $\alpha = 0.7725\pm 0.011$ and $\beta = 0.601\pm 0.024$.
By using Eq.\ref{age} with P = 12.33 days and the B-V = 1.131, we
estimate an 
age of 350 $\pm$ 50 Myrs for Kepler-210; this estimate appears reasonable given
its 
high degree of activity.

\subsection{Mean stellar density}

When a planet is transiting in front of its parent star,
we have the opportunity to accurately derive the ratio between
the stellar radius $R_{\star}$ and the orbital semimajor axis $a$ (cf.,
Table \ref{planetkeys}).
Combining this information with Kepler's third law, we can compute
an expression for the mean density $\rho _{mean}$ of the host star through
\begin{equation}
\label{rhomean}
\varrho _{\star,mean} = \frac{3 \pi} {G} \frac {a^3} {R^3_{\star} \ P^2},
\end{equation}
\noindent
with $G$ denoting the gravitational constant in addition to the terms containing
only {the observed quantities a/R$_{\star}$ and period P. The value for 
a/R$_{\star}$ is derived from the transit modeling, given the large number of
transits,
for both planets, the values of a/R$_{\star}$ and P can be estimated with
relatively
high accuracy.}
Carrying out this computation using the observed parameters for planets b and c
(cf., Table \ref{planetkeys}) we obtain densities of 
$\varrho_{\star,b}$ = 0.27 $\pm$ 0.004 $g/cm^3$ and $\varrho_{\star,c}$ = 0.46
$\pm$ 0.038 $g/cm^3$ 
, respectively for the host. 
On the other hand, based on the nominal stellar parameters of the host star we
expect a mean
density of $\varrho_o $ $\approx$   2.6 $g/cm^3$.  
Thus the mean host star densities 
derived from planets b and c are, first,  inconsistent with each other, and
second,
differ by almost an order of magnitude from the expected host star
density.
Since we firmly believe in Kepler's third law, there must be a physical
explanation for both discrepancies.

\subsubsection{Ellipticity of planetary orbits}
\label{sec:Elipt}

\begin{figure}[tp]
 \includegraphics[width=\columnwidth]{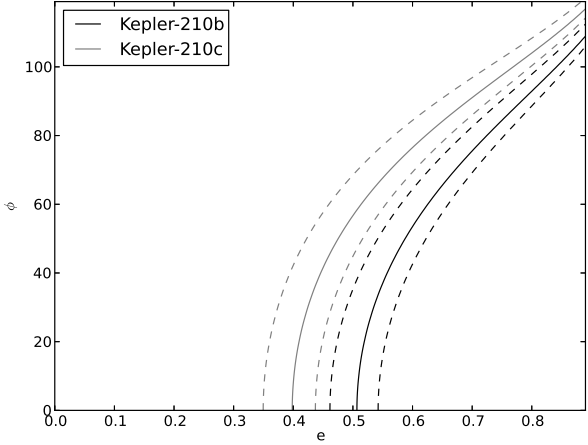}
\caption{Contour plot of eccentricity versus true anomaly during the 
mid-transit ($\phi$). 
For a circular system (e = 0) the density 
should equal $\varrho_o \; \backsimeq$  2.6 $g/cm^3$, as can 
be calculated for the given values of $R_{\star}$ and $M_{\star}$. 
The TBD
derived from the $a/R_{\star}$ values of $\varrho_{\star}$ for the
planets Kepler-210b (black line) and Kepler-210c {(gray line)} can be
explained for eccentricities $\gtrsim$ 0.4 and $\gtrsim$ 0.51 respectively,
depending
on the true anomaly of the planet during the mid-transit. The dashed lines 
represent the uncertainty limits.} 
\label{phidegen}
\end{figure}

\begin{figure}[bp]
 \includegraphics[width=\columnwidth]{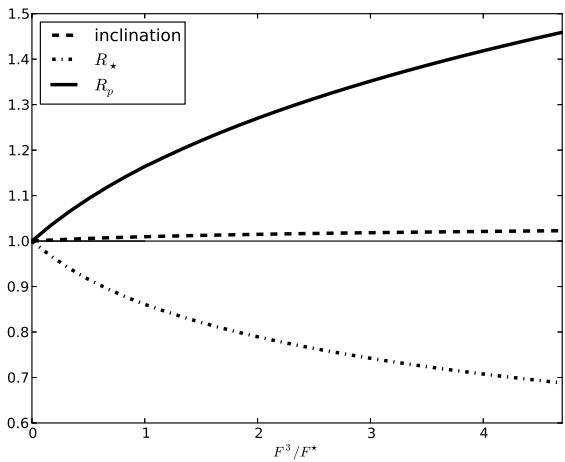}
 \caption{The variation of inclination, $R_{\star}$ and $R_{p}$, assuming third
light interference $F^3$ for Kepler-210c}
\label{newirsrp}
\end{figure}

So far our analysis has implicitly assumed circular orbits for both planets.
For elliptical orbits the orbital speed and hence the transit duration change
during the orbit and therefore there is no unique relation between transit
duration and stellar and planetary dimensions.   
Assuming that the orbital velocity
is constant during the actual transit, \cite{2005ApJ...627.1011T}
relate the transit duration $D_{ell}$ to the period $P$ and the impact parameter
$b$ of a transit through the expression

\begin{equation}
\label{dur}
D_{ell} = 
\frac{\sqrt{(1 - e^{2})}} {1 + e\cdot cos(\phi _t)}
\frac{P}{\pi}
\frac{\sqrt{(R_{\star}+R_{pl})^2 - b^2)}}{a},
\end{equation}
\noindent
where $\phi _t$ denotes the true anomaly at the mid-transit, 
while $R_{\star}$, $R_{pl}$ and $a$, denote stellar and planetary radii and
semi-major
axis respectively.  
Consequently, the transit duration $D_{ell}$ of an elliptical orbit scales with
the
transit duration $D_{circ}$ of a circular orbit (for the same system geometry
and the same period) through

\begin{equation}
\label{circel}
D_{ell} = 
\frac{\sqrt{(1 - e^{2})}} {1 + e\cdot cos(\phi _t)} \times D_{circ}
= g(e, cos(\phi _t)) \times D_{circ}.
\end{equation}
\noindent
It is straightforward to convince oneself that the derived sizes for star and
planet scale with the scaling function $g(e, cos(\phi _t))$ introduced in
Eq.\ref{circel}. Since the mean stellar density scales with $R^3_{\star}$, we
find

\begin{equation}
\label{mdenro}
\varrho_{\star ,ell}  = \varrho_{\star, circ} \cdot\left({\sqrt{1-e^2}} \over
{1+e
\cdot cos \phi _t} \right)^{-3}
\end{equation}
\noindent
with $\varrho_{\star, circ} = \varrho_{\star, mean}$ from Eq.\ref{rhomean}.
Hence the discrepant stellar densities can be explained by introducing
suitable eccentricities and true transit anomalies.  
By solving \ref{mdenro} for different values of $e$ and $\phi _t$ it is thus
possible to
to constrain the range of permissible eccentricities as well as values for $\phi
_t$, 
for which the derived stellar density becomes equal to the density expected
for the spectral type of the star for both planets; the corresponding curves are
shown in
Fig. \ref{phidegen}, where we plot for each planet the combination of
$e$ and $\phi _t$ resulting in a nominal stellar density of 2.7 $g\ cm^{-3}$.
Fig. \ref{phidegen} shows that eccentricities of 0.4 (for planet b) and 
0.5 (for planet c) are required to produce the expected stellar densities.

\subsubsection{Second (third) light scenario}

\begin{figure}[bp] 
 \includegraphics[width=\columnwidth]{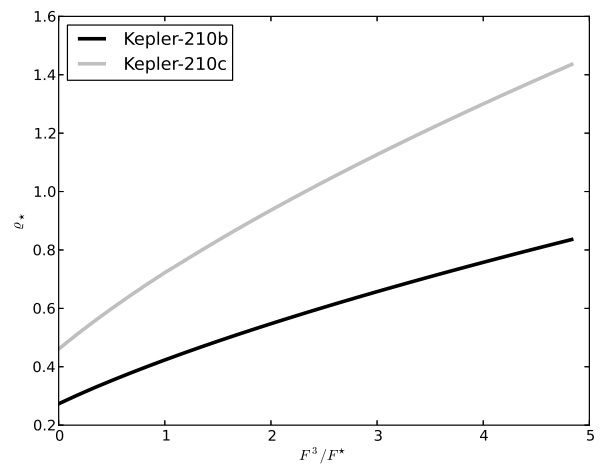}
 \caption{Derived stellar density versus assumed third light
contribution $F_3$}
\label{3rdl}
\end{figure}

Despite the fact that neither the optical spectrum nor the centroid analysis
of the {\it Kepler} data
have shown any evidence for a companion or blend, it might still be possible
that some {third} object in the background or foreground with flux $F_3$
contributes to the system flux in a way that the observed total flux $F_{obs}$
is given by

\begin{equation}
F_{obs} = F^\star + F_3 ,
\end{equation}

\noindent where $F^\star$ is the desired planet host's flux, which has to be
used for the
transit modeling.
If this {hypothetical} third light contribution $F_3$ is substantial, 
the true transit depth $d_{true}$ would be underestimated and 
{an incorrect radius for the Kepler-210 host star
would be derived}.
Assuming that the limb darkening coefficients are identical and
equal to the values presented in Table \ref{stellarkeys} for all system
sources, we calculate the influence of the third light source on the 
derived  stellar density (as shown in Fig.\ref{3rdl}), {considering
the following 
non-linear system of equations}, which allows computing the stellar and
planetary 
radii (each scaled by the semi-major axis) $\tilde{R}_\star$ and $\tilde{R}_p$
and, $i$, 
the inclination of the orbit normal with respect to the line of sight, given the
observed period
$P$, the observed time between the first and forth contact, $T_{14}$, the time
between
the second and third contact, $T_{23}$ and the observed (relative) transit depth
at mid-transit
$d_{obs}$.

\begin{eqnarray}
\label{seteq3rdl}
sin^{2}i\:cos^2 \left({{\pi}\over{P}}T_{14}\right) &=& 1-(\tilde{R}_\star +
\tilde{R}_p)^{2} \\ 
sin^{2}i\:cos^2 \left({{\pi}\over{P}}T_{23}\right) &=& 1-(\tilde{R}_\star -
\tilde{R}_p)^{2}
\end{eqnarray}
and
\begin{eqnarray}
\label{seteq3rdl2}
d_{true} = \frac{(1- c_1-c_2)+(c_1+2c_2)\:\mu_c-c_2\mu_c^{2}}{1-{c_1\over
3}-{c_2\over 6}}\left(\tilde{R}_p\over \tilde{R}_\star\right)^{2}
\end{eqnarray}
\noindent
Here $c_1$ and $c_2$ denote the quadratic limb darkening coefficients and 
$\mu_c$ denotes the expression
\begin{equation}
\mu_c = \sqrt{1-\frac{cos^{2}i}{\tilde{R}_\star^{2}}}.
\end{equation}
\noindent
We clearly need the true transit depth $d_{true}$ to compute
the values of $\tilde{R}_\star$, $\tilde{R}_p$ and $i$, yet
only the observed transit depth $d_{obs}$ is available; the two depths
are related, however, through 
\begin{equation}
d_{true} = (1 + d_{obs})\frac{F_3}{F^\star}.
\end{equation}
{Therefore, given the observed values of
$d_{obs}$, $T_{14}$ and $T_{23}$ and the observed periods $P$ for both planets},
the derived values for $\tilde{R}_\star$ and hence $\varrho_{\star }$ will
depend on the assumed third light contribution $F_3/F^{\star}$. 
{The resulting system of equation is quite non-linear.  In order to
provide a feeling on how sensitive the solutions depend on the third light
contribution $F_3/F^{\star}$, we plot in Fig. \ref{newirsrp} the variation of
the derived values for the inclination and stellar and planetary radii 
(for the planet Kepler-210c), 
relative to the case of no third light.  As is clear from Fig. \ref{newirsrp},
the inclination increases only slightly (it cannot exceed 90 degrees), while
the stellar radius decreases (as desired) and the planetary radius increases.
Finally, we can derive the stellar density, for which}
our results are shown in Fig. \ref{3rdl},
where we plot the derived stellar densities for both planets as a function of
the assumed third light contribution  $\frac{F_3}{F^\star}$.
As is clear from Fig. \ref{3rdl}, the third light contribution would have to be
substantial, and in fact
{the third light would have to dominate the total system flux
in order to obtain values of  $\varrho_{\star c}$ as expected for stars on the
main sequence in the relevant spectral range.}
Yet, the two planets still yield discrepant densities of their host,
so one would have to introduce yet another host for the second planet, which
appears at least a little contrived.  Therefore we conclude that the
introduction of a third light source does not lead to a satisfactory solution
of inconsistency in the derived stellar parameters.

\subsubsection{Inflated star}

Another possible scenario explaining the {\it Kepler} observations of
Kepler-210
would be the assumption that the host is not on the main sequence, but rather
evolved and in fact a giant or sub-giant. Such stars are usually not active,
however, there are some classes of evolved stars which are quite active, for
example,
variables of the FK Com type. Those stars are highly active G-K type sub-giant 
stars with surface gravities log(g) of $\sim$ 3.5. They show strong 
photometric rotational modulations caused by a photosphere
covered with inhomogeneously distributed spots. An 
other important characteristic of these objects is their rapid rotation.
Generally
the  $v\:sini$ derived from their spectra is between $\sim$
50 and 150 $km\cdot s^{-1}$ \citep{2005LRSP....2....8B}. In the case of
Kepler-210
$v\:sini$ is  $\sim$ 4 $km\cdot s^{-1}$, and thus
we believe that a FK-com scenario does not provide
a suitable explanation for the observed density discrepancy. 

\subsection{Kepler-210 TTVs}

\hspace{3 ex} As described in sec. \ref{sec:ttvs}, TTVs are detected in 
both planets. 
In order to further examine the properties of these variations we searched for
any periodicities in the O-C data by constructing a Lomb-Scargle periodogram on
every
set. For the outer and larger planet, the Lomb-Scargle periodogram shows a
leading period 
of about 690 days, which is apparent in the modulation of the O-C curve in
Fig.\ref{ocdiag}, 
while for the inner and smaller planet the periodicity results 
remain ambiguous, most probably due to the large scatter in its O-C diagram.

\begin{table}[bp]
\begin{center}
\caption{$\chi^2$ results of the model for 2 eccentric planets
hypothesis, 3 non eccentric planets and 3 eccentric planets. In the square
brackets are listed the degrees of freedom.}\label{ttvchisqr}
\begin{tabular}{rrrr}
\hline
\hline
\\[-8pt]
  & 2 ecc Planets & 3 non ecc Planets & 3 ecc Planets\\
\hline
\\[-6pt]
b & 489.57 [420] & 447.9 [415] & 443.9 [415] \\
c & 116.22 [86] & 112.4 [81] & 117.59 [81]  \\
\hline
\end{tabular}
\end{center}
\end{table}

What would be a physical scenario consistent with these O-C diagrams?
We first note that the orbital period ratio of the system is very
close to a 13/4, if we consider that the errors in periods in Table
\ref{planetkeys} are also affected by the TTVs. 
This ratio is not close to any low order mean motion resonance so the
amplitude of any TTVs is expected to be relatively small for both planets
\citep{2005MNRAS.359..567A}.
 
In order to verify this and to model the TTVs we use the
N-body code as presented in the same paper. 
The N-body code requires the planetary masses which are unknown.
{Assuming ad hoc that the planetary densities are below
5 g/cm$^3$, it is clear that the masses of the two planets are 
substantially below 0.5 M$_J$. Furthermore,
in order to roughly estimate the planetary masses below that limit,} 
we use a general mass vs. radius law as described 
by \cite{2011ApJS..197....8L} in the form
\begin{eqnarray}\label{eqmasrad}
M_{p} = R_{p}^{2.06}
\end{eqnarray}
\begin{figure}[tp]
 \includegraphics[width=\columnwidth]{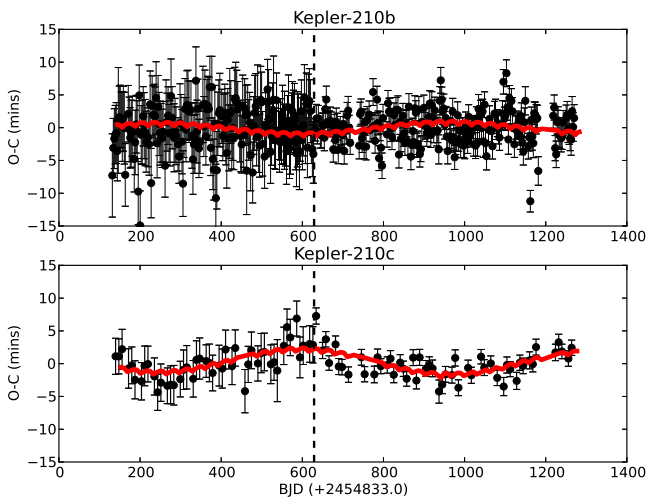}
 \caption{TTVs expected for two planets assuming eccentric orbits
{with e$_b$ = 0.44 and e$_c$ = 0.50.}}
\label{2ecc}
\end{figure}
\noindent
With the masses thus specified, we first considered only the two 
transiting planets for our TTVs simulations.
Assuming non-eccentric orbits resulted in TTVs of less than a minute, which is
far from the observed variations for both planets.  The TTVs would
remain in that state even if we assume higher masses, under the limit of 12
M$_{J}$. As discussed in detail by \cite{2012ApJ...761..122L}, the TTVs
amplitude can also be affected by eccentricity. 
Implementation of eccentric orbits for both planets, Kepler-210b and
Kepler-210c, improved the fit substantially; the modeled TTVs together with the
data are
shown in
Fig.\ref{2ecc}, the fit results in terms of fit quality measured through $\chi
^2$ are given in Table \ref{ttvchisqr}.

Clearly, also the TTVs analysis supports a scenario of two planets with rather
eccentric orbits similar to our discussion in \ref{sec:Elipt}.  
However, in order to produce the observed TTVs the system configuration must be
such 
that the true anomaly, of both planets at the time of transit, $\phi_{t}$,
should exceed 40$^{\circ}$,  while also the difference in true anomaly
$\Delta\phi_{t}$ should be $\sim$
60$^{\circ}$, However, this configuration appears impossible due to the
physical constraints in Fig.\ref{phidegen}.  Furthermore, our stability
tests, (which performed with the ${\bf swift\_rmvs3}$ algorithm 
\citep{1994Icar..108...18L}, show
that this configuration is unstable on time scales in excess of $\sim$ 1 Myear. 
{Finally we note that the probability of observing a transit is
higher for small values of true anomaly, i.e., for the times near periastron
passage.}  

We therefore conclude that a scenario with only two planets with eccentric
orbits is unlikely and introduce a third, hypothetical planet KOI-676.03 in
order to stabilize the system.
{We consider both eccentric and non-eccentric orbits for Kepler-210b and 
Kepler-210c. 
In order to determine period, mass and eccentricity for the hypothetical 
planet KOI-676.03, we considered several possible system configurations.}

{We emphasize that we cannot derive a unique solution for the
physical parameters of this hypothetical planet.  Most importantly,
we need to assume a mass for this planet, which controls
the strength of the gravitational interaction with the observed planets
Kepler-210b and Kepler-210c.  Thus, given the observed TTVs amplitude and
given the assumed mass of KOI-676.03, a certain value of semi-major 
axis and hence period is derived.  The higher the assumed mass,
the longer the resulting period, and thus there
is more than one configuration to account for the detected TTVs
signal.}

\begin{figure}[tp]
 \includegraphics[width=\columnwidth]{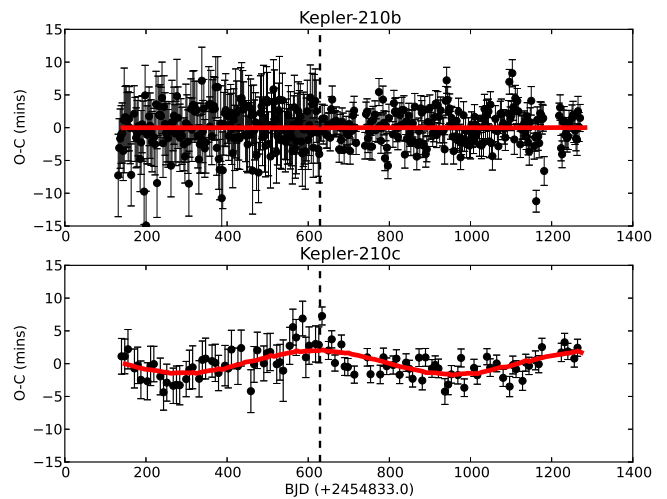}
 \caption{TTVs expected for three planets with non-eccentric orbits
{for all components of the system. The third planet's period for that
case is P$_{03}$= 63.07 days.}}
\label{3noecc}
\end{figure}

\begin{figure}[tp]
 \includegraphics[width=\columnwidth]{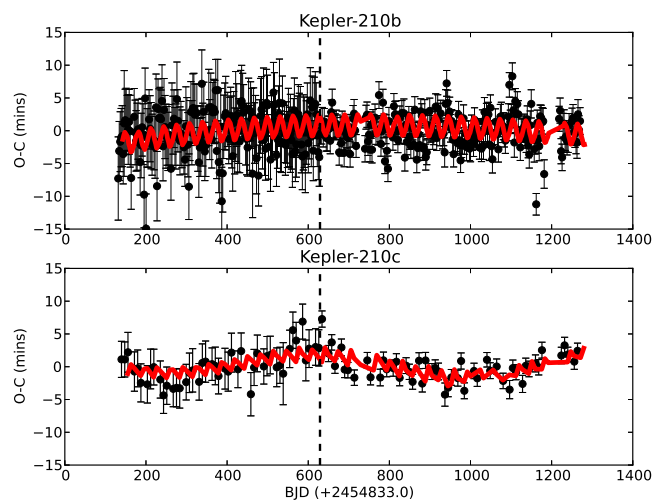}
 \caption{TTVs expected for three planets assuming eccentric orbits
{with e$_b$ = 0.45, e$_c$ = 0.51 and e$_{03}$ = 0.23. The Third planet's
period for that case is
P$_{03}$ = 63.29 days.}}
\label{3ecc}
\end{figure}

{In order to produce possible candidate systems, we carried
out simulations assuming some given mass for KOI-676.03, considering
periods between 20 to 300 days, masses in the range $M_{03} \sim$ 
0.1-0.6 M$_J$ and eccentricities e $\simeq$ 0.1-0.3. A particularly
promising configuration, but by no means unique solution, consistent with all
{\it Kepler} data, has
a period P $\simeq$ 63 days;} in Fig.\ref{3noecc} and again 
Table \ref{ttvchisqr} (for the case with zero eccentricity) and 
in Fig.\ref{3ecc} and Table \ref{ttvchisqr} (for the
eccentric case) we show that such a scenario provides results consistent with
the
available {\it Kepler} data.
As is clear from Fig.\ref{3noecc} and Fig.\ref{3ecc}, as well as
Table \ref{ttvchisqr}, the
difference between the non-eccentric and eccentric case is marginal 
at best, while (statistically) preferable over a two planets scenario.
In addition, the eccentricities of the planets Kepler-210b and Kepler-210c give
a high frequency TTVs signal, which might better explain the higher
dispersion of the TTVs in Kepler-210b. In that case the model also suggests
$\phi_t$ values around zero with $\Delta\phi <$ 30 $^\circ$, which are in line
with Fig.\ref{phidegen}. Also the system's stability exceeds 10$^7$ years. 

\begin{figure}[tp]. 
 \includegraphics[width=\columnwidth]{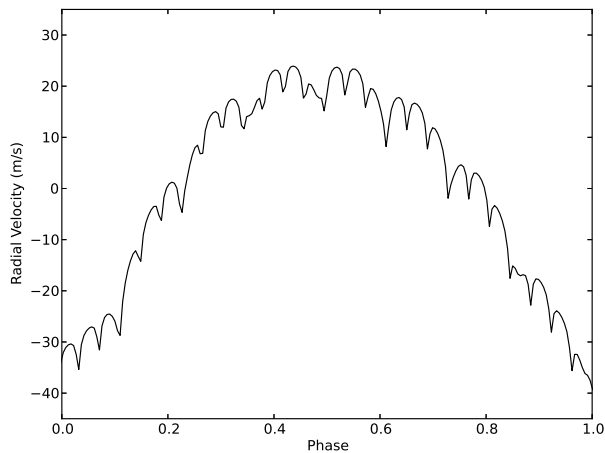}
 \caption{The predicted, synthetic, radial velocity diagram for the
system Kepler-210 for P$_{03}$ = 63 days, M$_{03}$ = 0.4 M$_j$, e$_{03}$ = 0.23
and i$_{03}$ = 60$^\circ$.}
\label{rv}
\end{figure}

In the case of non eccentric orbits the system would reach fatal instability
once
masses above 5 M$_J$ are chosen. For eccentric orbits the upper limit for the
masses of the system becomes lower. While this fact suggests
a planetary nature of the system components, it also introduces an
additional factor of concern about the long term stability of the system.
We do point out that this third stabilizing planet does produce a radial
velocity signal in the system.  
For our nominal case we plot in Fig.~\ref{rv} the expected RV signal in a
synthetic radial velocity diagram, which shows peak-to-peak variations in excess
of 
60 m/sec;  clearly such RV variations ought to be detectable despite the high
activity level of the host star, and therefore, the detection of a RV-signal
would significantly constrain the possible configuration space of the system.

\section{Summary}
\begin{figure}[bp]
 \includegraphics[width=\columnwidth]{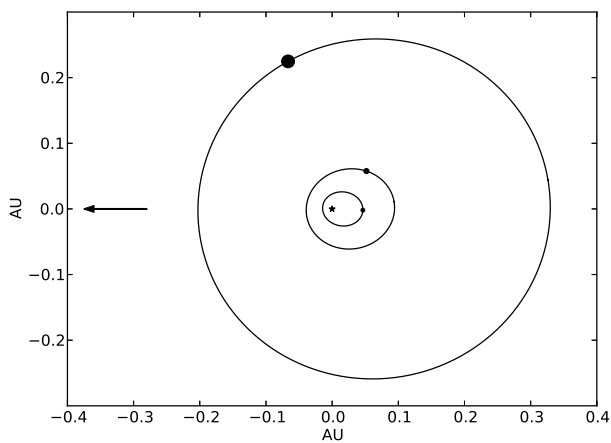}
 \caption{The suggested system configuration. The earth is at the
direction of the arrow.}
\label{}
\end{figure}

We report the detection and characterization of two transiting Neptune-sized
planets, Kepler-210b and Kepler-210c with periods of 2.4532
days and 7.9723 days respectively around a presumably quite young and active
K-type dwarf.  These objects were first listed as planetary candidates by 
the {\it Kepler} team. We show that the transits of both planets are
affected by spots.  From the observed transit parameters and in particular from
the
observed value of $a/R_{\star}$ it is possible to calculate the mean density of
the host star using the Kepler's 3$^{rd}$ law.  
Interestingly, the two planets yield discrepant
mean host star densities, which in addition are inconsistent with the
densities expected for a K-type dwarf.  Having explored various
scenarios we conclude that the assumption of quite eccentric orbits for both
planets provides
the currently most probable scenario.

In addition, both planets show transit timing variations. 
Using N-body simulations we constructed possible system configurations
consistent with the {\it Kepler} data.  While it is possible to explain the
observed
TTVs with a two planet scenario, such a scenario requires a very special
geometrical 
configuration and is unstable on time scales of 10$^6$ years.
As a result we believe that there exists a third planet, KOI-676.03,
with a mass between $\sim$ 0.3-0.6 M$_J$ in a 
slightly eccentric (e $\simeq$ 0.2) orbit with period $\sim$ 63 days, which 
stabilizes the whole Kepler-210 planetary system. 

{Different configurations are also possible,} 
yet this stabilizing hypothetical planet should produce a detectable RV
signal. 
We therefore suggest RV monitoring of Kepler-210, which is likely to provide a
substantially increased insight into the Kepler-210 planetary system.

\section{Acknowledgments}

PI and CvE acknowledge funding through the DFG grant RTG
1351/2 ''Extrasolar planets and their host 
stars''.  The results of this publication are based on data collected by the NASA's
{\it Kepler} satellite. 
We would like to thank the director and the Calar
Alto Observatory staff for the approval and execution of a DDT proposal.
Finally we acknowledge the exceptional work by the CFOP team which was
essential for this publication.

\bibliographystyle{aa}
\bibliography{references}

\end{document}